# Stimulated excitation of resonant Cherenkov radiation at a large number of neighbouring waveguide modes

L. Sh. Grigoryan[1], S. R. Arzumanyan, H. F. Khachatryan, M. L. Grigoryan

*NAS Institute of Applied Problems in Physics, Yerevan Armenia*

**Abstract.** The resonance Cherenkov radiation generated from a train of equally-spaced unidimensional electron bunches travelling along the axis of a hollow channel inside an infinite cylindrical waveguide filled with (weakly dispersing) transparent dielectric has been investigated. It was shown that its excitation might be stimulated at a large number of neighboring modes of the waveguide. A visual explanation of this effect is given and the possibility of its observation in the range of terahertz radiation is discussed.

**Keywords:** Waveguide, train of bunches, resonant Cherenkov radiation, stimulated excitation of radiation.

## 1. Introduction

With the advent of modern accelerators it became possible to generate small size ($<1mm$) relativistic monoenergetic bunches with large number of electrons ($>10^9$) to produce high power quasi-coherent terahertz radiation that is of considerable interest for applications in physics, chemistry and biology [1]. Direct observation of the record power, narrow-band and coherent terahertz Cherenkov radiation (CR) generated by such a bunch travelling through a hollow channel inside the dielectric-filled waveguide was reported in [2] (2009). As was clarified later on (2010–2011), the radiated power of CR attained in [2] could have been essentially increased, if the filler dielectric in the waveguide were cut into plates and spaced at some distances one from the other [3,4].

There is another theoretical opportunity for considerable increase of the power of CR attained in [2] that is connected with the fact that in accelerators it turns possible to generate the trains of bunches spaced at small distances $d$ ($<1mm$). Based on this fact high power resonant [6] (the so-called parametric [7]) coherent THz CR has been generated in [5] (2011) at a separate waveguide mode $\omega_{s_0}$ by selecting $d$ distance according to the well-known formula

$$d = 2\pi v / \omega_{s_0} \qquad (1a)$$

( v is the velocity of bunch train travel and $s_0$ is the number of emitted mode).

Another choice of the value of $d$ suggested in [8] (2006) provided the generation of resonant CR at larger number of neighboring modes of the waveguide, that proved to be higher power than in the case of (1a). Ibidem a visual explanation of this effect has been also given. However, the study in [8] was concerned with the case of CR from *a train of point charges* travelling along the axis of *a waveguide completely filled with a dielectric*. In reality (i) a train of electron bunches, not that of separate electrons is generated and (ii) the ionization losses are diminished when this train is directed in the hollow channel inside the dielectric. The presence of a hollow channel undoubtedly reduces the power of CR (see [9]) and complicates theoretical analysis of the problem. In the present work both these facts have been taken into account to construct a more adequate model of the situation.

---
[1] E-mail address: levonshg@mail.ru

## 2. The problem formulation

Let us consider a train of equidistant unidimensional electronic bunches traveling along the axis of hollow channel in the central part of dielectric-lined waveguide[2]. In case of electron velocity

$$v > c/\sqrt{\varepsilon\mu} \qquad (2)$$

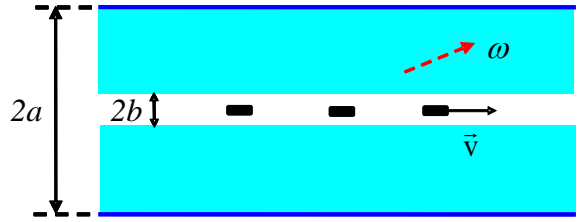

Fig.1. A train of three equidistant unidimensional electronic bunches traveling along the axis of hollow channel in the central part of dielectric-lined waveguide.

($\varepsilon$ and $\mu$ are the permittivity and permeability of the substance of waveguide filler respectively) the CR is generated inside the waveguide at discrete frequencies $\omega_s$ determined from the following dispersion equation

$$k\varepsilon\varphi_0 I_0(kb) = \chi\varphi_1 I_1(kb) \qquad (3)$$

(see, e.g., [9]), where

$$\varphi_0 = J_1(\chi b)N_0(\chi a) - J_0(\chi a)N_1(\chi b)$$
$$\varphi_1 = J_0(\chi b)N_0(\chi a) - J_0(\chi a)N_0(\chi b) \qquad (4)$$
$$k = \frac{\omega_s}{v}\sqrt{1-v^2/c^2}, \quad \chi = \frac{\omega_s}{v}\sqrt{\varepsilon\mu v^2/c^2 - 1}.$$

Here $a$ and $b$ are respectively the radii of waveguide and of the hollow channel in its center (Fig.1), $J_n(x)$ and $N_n(x)$ are the Bessel functions of the first and second kind of order $n$, and $I_n(x)$ is the modified Bessel function of the first kind of order $n$.

The energy of CR generated in unit time is determined by the sum

$$\sum_{s=1}^{\infty} P(s) = \sum_{s=1}^{\infty} P_1(s)\,F(s), \qquad (5)$$

where [9]

$$P_1 = 2\frac{q^2}{v}(1-v^2/c^2)\frac{\omega_s[\chi\varphi_1 K_1(kb) + k\varepsilon\varphi_0 K_0(kb)]}{d[\chi\varphi_1 I_1(kb) - k\varepsilon\varphi_0 I_0(kb)]/d\omega_s} \qquad (6)$$

is the radiation power of a single electron at the $s$-th waveguide mode ($K_n(x)$ is the modified Bessel function of the second kind of order $n$). In (5)

$$F = n_q(1-f_q)n_b + n_q^2 f_q n_b^2 f_{tr} \qquad (7a)$$

---
[2] Hereinafter we shall assume that the waveguide is filled with a transparent substance.

is the structural factor [6,10-12] of the train of bunches. Here $n_q$ is the number of electrons inside a bunch,

$$f_q = \exp(-\omega_s^2 \sigma^2 / v^2) \qquad (7b)$$

is the coherence factor of CR from electrons in a single bunch (admittedly distributed according to Caussian law with standard deviation $\sigma$), $n_b$ is the number of bunches, and

$$f_{tr} = \sin^2(\omega_s d n_b / 2v) / n_b^2 \sin^2(\omega_s d / 2v) \qquad (7c)$$

is the coherence factor of radiation from bunches in the train.

According to (7c) the train of bunches coherently generates parametric [7] (or, so called, resonant [6]) CR at the separately taken mode $s_0$: $f_{tr}(s_0) = 1$, if

$$d = 2\pi m v / \omega_{s_0}, \qquad \text{where} \qquad m = 1;2;3.... \qquad (1b)$$

*The novelty* of *our work* consists in the other choice of the distance between bunches:

$$d \cong m d_0, \qquad \text{where} \qquad d_0 = 2(a-b)\sqrt{\varepsilon \mu v^2 / c^2 - 1} \qquad (8)$$

(and in a visual derivation of the above equality) in case of which the CR from the train of bunches is practically resonant at a large number of neighboring modes:

$$f_{tr}(s) \approx 1, \qquad s_1 \leq s \leq s_2, \qquad (s_2 - s_1) \gg 1. \qquad (9)$$

### 3. Numerical results

In our calculations we have used (3) – (8), $b = 0.1a$, and

$$d = 2(1+\alpha)d_0, \qquad \text{where} \qquad \alpha = 0.011. \qquad (10)$$

The energy of electrons was taken to be $10 MeV$, $n_q e = 200 pC$, $\varepsilon = 3.8$, $\mu = 1$ [2]; $\sigma = 0.02a$ and $n_b = 5$.

The values of $P(s)/P_*$ are shown by blue dots in the lower part of Fig.2, where $P_* = 10v(en_q n_b)^2 / a^2$. On comparing the values of $f_{tr}(s)$ and $f_q(s)$ in the upper part of Fig.2 one makes certain that

$$f_{tr}(s) \approx 1 \quad \text{and} \quad f_q(s) \geq 0.5 \quad \text{at} \quad 10 \leq s \leq 20. \tag{11}$$

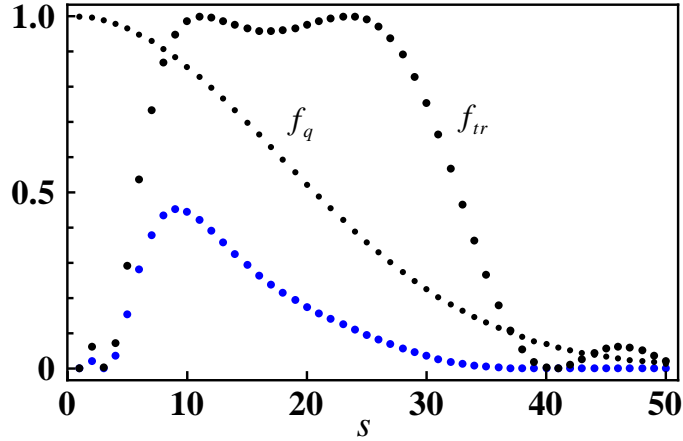

Fig.2. With blue points the power $P(s)/P_*$ of CR from a train of $n_b = 5$ unidimensional electron bunches was plotted depending on the number $s$ of the radiated waveguide mode, $P_* = 10v(en_q n_b)^2 / a^2$. The calculations were made using formulae (3)-(8). Black points represent the coherence factors $f_q(s)$ (for the radiation generated by a separate bunch) and $f_{tr}(s)$ (for the radiation generated by the train as a whole).

In this relatively wide range of waveguide modes *the energy of CR from the train of bunches is* $\approx n_b^2 = 25$ *times as large as that from the separate bunch* (see (5),(7a),(9)). Here in case of $a = 2mm$ the power of CR is

$$\sum_{s=6}^{24} P(s) = 36.3 MW. \tag{12}$$

The radiation is generated within the range of $273 \leq \omega_s / 2\pi \leq 960\, GHz$.

### 4. Visual explanation

Equality (8) admits a visual interpretation. Now let us consider the instant when the first bunch (the rightmost one in the direction of train travel) was in the vicinity of point $A$ (see Fig.3a) and observing the laws of ray optics follow the propagation of (the dotted) CR pulse generated by this bunch in the vicinity of point $A$. The angle $\theta$ in Fig.3a specifies the direction of radiation propagation in an infinite homogeneous medium: $\cos\theta = c / v\sqrt{\varepsilon\mu}$. At some moment of its propagation the CR pulse would come closest to the trajectory of bunches (the waveguide axis) in the vicinity of point $C$. By this time the pulse would lag behind its faster-than-light source which will be found rightwards in the vicinity of $A_C$ point. The second bunch stays leftwards in the vicinity of point $B$ at the distance $d$ from the first one (see Fig.3b, in Fig.3a it is not shown), and if

$$CB = 0 \quad \Leftrightarrow \quad d = d_0 \tag{13}$$

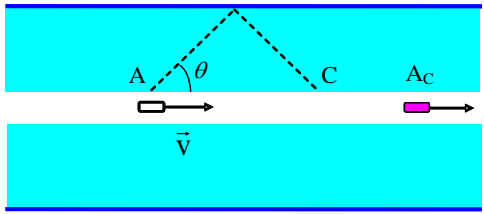
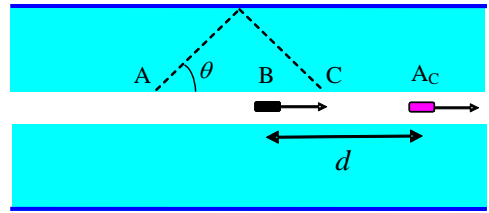

Fig.3a                                                                Fig3.b

Fig.3a. In the vicinity of point $C$ the propagating Cherenkov wave (dotted line), emitted by the first bunch of train in the vicinity of point $A$, would most closely approach the trajectory of bunches. By this moment of time the faster-than-light bunch will be found rightwards, in the vicinity of point $A_C$.

Fig.3b. At the same moment the second bunch of the train (it is not seen in Fig.3a) will be located leftwards, - in the vicinity of point $B$, at $d$ distance from the first one.

then it will traverse the vicinity of point $C$ simultaneously with the pulse. In this case *an exceptional situation* is realized when the CR pulse emitted by the second bunch in point $B$ is formed and simultaneously superimpose on the pulse that was emitted earlier by the first bunch in flight in the vicinity of point $A$. The situation with all $d$ that are multiples of $d_0$:

$$d = md_0 \qquad (14)$$

is similar.

One may conclude by analogy that in areas immediately adjacent to each of bunches (except for the first one), i.e., in the radiation formation zone the processes of emission and superposition of pulses occur simultaneously. In case of weak dispersion the condition (14) is practically independent of the number of generated waveguide mode and, hence, *the superposition of pulses will occur on the larger number of neighboring modes*. As a result of interference the pulses may be either suppressed or amplified depending on the specific value of phase difference. The in-phase superposition of electromagnetic oscillations would go with an amplification of field in the zone of radiation formation. As a result, the force retarding the motion of bunches also increases, and an additional work performed by the external force constraining the uniform motion of bunches shall be used for formation of higher power resonant CR stimulated on the larger number of neighboring modes. This case is illustrated in Fig.2 for $m = 2$.

The second equality in (13) is valid in the frameworks of geometrical optics laws, when the wavelength of the emitted wave is much less than $b$. This method for determination of $d$ requires a more accurate definition for modes with $s \leq a/b$. It is not surprising, hence, that in case of $a/b = 10$ the resonant CR, the excitation of which is stimulated on modes with $6 \leq s \leq 20$ (see Fig.2), occurs for $d$ that somewhat differs from (14), namely for $d$ that is determined according to (10).

**5. Discussion**

The coherent terahertz CR generated by a subpicosecond bunch of relativistic electrons travelling along the axis of a waveguide filled with a hollow dielectric has been experimentally studied in [2]. Here the value of $b/a \approx 0.23$ was chosen so that CR were generated only on the first waveguide mode. Later on it became possible [5] to generate the resonant (and coherent) THz radiation of a train of subpicosecond electronic bunches on a separate mode of the waveguide by proper choice of $d$ distance by formula (1a).

We propose another choice of $d$ (see (10)) providing an emission of resonant (and coherent) THz radiation, the excitation of which is stimulated on the larger number of neighboring waveguide modes. An increase in the number of resonantly emitted modes is stipulated by the fact that in areas immediately adjacent to each bunch (except for the first one) both the processes, - the emission and constructive interference of CR waves emitted earlier by those bunches flying ahead, occur simultaneously.

Similar stimulation for transition radiation was observed experimentally in [13-15]. In these works the total energy of radiation has been measured (confer with (12)) in the far infrared range. The radiation was generated by a train of electron bunches at the SUNSHINE setup (Stanford University, USA). The observed stimulation effect of transition radiation was explained by the interference of waves nearby the radiator, i.e., in area immediately adjacent to the emitting bunches. For implementation of such an interference a special system of mirrors and reflectors was constructed and adjusted to direct the transition radiation from one bunch of electrons to the other (confer with Fig.3b).

## 6. Conclusions

1) In the present work the Cherenkov radiation from a train of equally-spaced unidimensional electron bunches travelling along the axis of a hollow channel inside an infinite cylindrical waveguide filled with (weakly dispersing) transparent dielectric has been investigated.
2) It was shown that if the distance between the bunches is determined by equality (10), then the excitation of resonant CR from a train of bunches may be stimulated on the larger number of neighboring waveguide modes (Fig.2).
3) A visual explanation of this effect was given (Section 4) based on the fact that in case of weak dispersion the equality (10) is practically independent of the number of generated waveguide mode.
4) It is proposed to use the effect of stimulated excitation of the resonant CR on the larger number of neighboring modes for an additional increase of the power of experimentally observed resonant and coherent THz radiation [2,5].

## Acknowledgments
One of the authors (L.Sh.G.) is thankful to B.M. Bolotovskii for stimulating discussions, P.A. Alexandrov, N.F. Shulga, S.I. Zaicev and L.A. Gevorgian for valuable comments.